\documentclass[twocolumn,prl,aps,epsfig]{revtex4}
\usepackage{psfig}

\def\beq{\begin{equation}}
\def\eeq{\end{equation}}
\def\beqa{\begin{eqnarray}}
\def\eeqa{\end{eqnarray}}
\def\MeV{\nobreak\,\mbox{MeV}}
\def\GeV{\nobreak\,\mbox{GeV}}

\def\md{m_D}

\def\mpsi{m_\psi}

\newcommand{\fslash}[1]{\ooalign{\hfil/\hfil\crcr$#1$}}

\begin{document}

\title{$J/\psi$ dissociation by pions in QCD}

\author{Francisco O. Dur\~aes$^1$, Su Houng Lee$^{2,3}$, Fernando S. 
Navarra$^1$ and Marina Nielsen$^1$}
\affiliation{$^1$Instituto de F\'{\i}sica, Universidade de S\~{a}o Paulo, 
C.P. 66318,  05315-970 S\~{a}o Paulo, SP, Brazil\\
$^2$Institute of Physics and Applied Physics, Yonsei University,
Seoul 120-749, Korea\\
$^3$Cyclotron Institute, Texas A\&M University, College Station, 
TX 77843-3366, USA.}

\begin{abstract}
We compute the $J/\psi~\pi\rightarrow \mbox{charmed mesons}$ cross section 
using QCD sum rules. This cross section is important to distinguish the
suppression in the production of $J/\psi$ through the dissociation by
comoving pions and through the formation of quark-gluon plasma. Our sum rules
for the $J/\psi~\pi\rightarrow \bar{D}~D^*$, $D~\bar{D}^*$, 
${\bar D}^*~D^*$ and ${\bar D}~D$ hadronic matrix elements are constructed by 
using vacuum-pion correlation functions, and we work up to twist-4.
After doing a thermal average we get $\langle\sigma^{\pi J/\psi} v\rangle\sim
0.2 - 0.4$ mb at $T=150\MeV$.
\end{abstract}

\maketitle

\vspace{1cm}

In  relativistic  heavy  ion  collisions $J/\psi$ suppression has
been recognized as an important tool  to  identify  the  possible
phase  transition  to  quark-gluon  plasma (QGP). Matsui and Satz
\cite{ma86} predicted that in  presence  of  quark-gluon  plasma,
binding  of  a  $c\bar{c}$  pair  into  a  $J/\psi$ meson will be
hindered, leading to the so called $J/\psi$ suppression in  heavy
ion  collisions.  Over the years several experiments measured the
$J/\psi$ yield in heavy ion collisions (for a review of data  and
interpretations see Ref.\cite{vo99,ge99}). In brief, experimental
data  do  show  suppression \cite{NA50}. However, this could be attributed to
more  conventional $J/\psi$  absorption by comovers, not  present   in   pA
collisions  \cite{kha,wong,cap}.  In order to confirm that the suppression
of $J/\psi$ comes from the presence of the QGP, it is necessary to understand
better the $J/\psi$ dissociation mechanism by collision with
comoving hadrons. 

Since there is no empirical information on $J/\psi$ absorption cross sections
by hadrons, theoretical models are needed to estimate their values.
In general, different models apply to different energy regimes and one of the
first estimates of the charmonium-hadron cross section uses short distance
QCD \cite{bhp,kha2,agga}. However, the method is inapplicable at low energies, 
which is
the regime of greatest interest for $J/\psi$ collision with
comoving hadrons. Besides, even in the high energy regime, nonperturbative
effects may be important~\cite{dnn} and can increase significantly the value of
the cross section. At the low energy regime one can use quark-interchange
models \cite{qmodel} or meson exchange models 
\cite{mmodel,nnr}. The results of the calculations for the 
charmonium-pion cross section based on these two approaches can differ by 
two orders of magnitude in the relevant energy range. Moreover, 
the rapid increase of the cross section near the threshold, obtained with 
these two models, is probably overestimated since these models do not respect 
chiral symmetry, as showed in \cite{nnr}. There is also a calculation of
the $J/\psi-\pi$ cross section \cite{nnmk} based on the QCD sum rules (QCDSR) 
technique~\cite{svz,rry}, which is also valid at the low energy regime. The 
result for the cross section in ref.~\cite{nnmk} is  in between the results 
in the quark-interchange models and meson exchange models.

In this work we improve the calculation done in ref.~\cite{nnmk} by considering
sum rules based on a three-point function with a pion. We work up to twist-4,
which allows us to study the convergence of the OPE expansion. Since
the method of the QCDSR uses QCD explicitly, we believe that our work 
will improve the understanding of this important topic.

In the QCDSR approach, the short range perturbative QCD is extended by an 
OPE expansion of the correlator, giving a series in inverse powers of
the squared momentum with Wilson coefficients. The convergence at low
momentum is improved by using a Borel transform. The coefficients involve 
universal quark and gluon condensates. The quark-based calculation of
a given correlator is equated to the same correlator, calculated using 
hadronic degrees of freedom via a dispersion relation, giving sum rules
from which a hadronic quantity can be estimated. 

Let us start with the vacuum-pion correlation function for the process 
$J/\psi~\pi\rightarrow \bar{D}~D$:
\beqa
\Pi_{\mu} &=& \int d^4x~d^4y~e^{-ip_2.y}~e^{ip_3.x}~
\nonumber\\
&\times&
\langle 0|T\{j_D(x)j_{\bar{D}}(0)j_\mu^\psi(y)\}|\pi(p_1)\rangle \;, 
\label{cor}
\eeqa
with the currents given by  $j_\mu^\psi=\overline{c} \gamma_\mu c$,
$j_D=\overline{u}i\gamma_5 c $ and
$j_{\bar{D}}=\overline{c} i\gamma_5d$.
$p_1$, $p_2$, $p_3$ and $p_4$ are the four-momenta of the 
mesons $\pi$, $J/\psi$, $D$ and $\bar{D}$ respectively. The advantage of 
this approach as compared with the 
4-point calculation in ref.~\cite{nnmk}, is that we can consider more terms
in the OPE expansion of the correlation function in Eq.~(\ref{cor}) and,
therefore, we get a much richer sum rule.

Following ref.~\cite{kl}, we can rewrite Eq.~(\ref{cor}) as:
\beqa
\Pi_{\mu}&=&\int {d^4k\over(2\pi)^4}~Tr[S(p_3-k)\gamma_\mu 
\nonumber\\
&\times&
S(p_3-p_2-k)\gamma_5 D_{aa}(k,p_1)\gamma_5]\;,
\label{corqq}
\eeqa 
where $S(p)$ is the free $c$-quark propagator, and $D_{ab}(k,p)$ denotes the
quark-antiquark component with a pion, which can be separated into three 
pieces depending on the Dirac matrices involved \cite{kl}:
\beq
 D_{ab} (k,p) = \delta_{ab} \left [ i\gamma_5 A+
\gamma_\alpha \gamma_5 B^\alpha 
+\gamma_5 \sigma_{\alpha \beta} C^{\alpha \beta}\right]\ .
\eeq
The three invariant functions of $(k,p)$: $A,\,B^\alpha$ and 
$C^{\alpha \beta}$, are 
defined by the Fourier transform of the vacuum-pion matrix elements:
$\langle0|\bar{d}(x)i\gamma_5 u(0)|\pi(p_1)\rangle$, $\langle0|\bar{d}(x)
\gamma^\alpha\gamma_5 u(0)|\pi(p_1)\rangle$ and  $\langle0|\bar{d}(x)
\sigma^{\alpha\beta}\gamma_5 u(0)|\pi(p_1)\rangle$ respectively.

Using PCAC and working at the order 
${\cal O}(p_{\mu} p_{\nu})$ we get up to twist-4 \cite{kl,bbk}:
\beqa
A(k,p)&=&{(2\pi)^4\over12}{\langle {\bar q} q \rangle \over f_\pi}\left[
-2+ip_{\alpha_1}{ \partial\over  i\partial k_{\alpha_1}}
+{1\over2}
\left(-{m_0^2\over4}\right.\right.
\nonumber\\
&\times&\left.\left. g_{\alpha_1 \alpha_2}
+{2 \over 3}p_{\alpha_1} p_{\alpha_2}\right){ \partial\over  i\partial 
k_{\alpha_1}}{ \partial\over  i\partial k_{\alpha_2}}\right]
\delta^{(4)}(k)\;,
\nonumber\\
B_\alpha(k,p)&=&{(2\pi)^4\over12}f_\pi\left[ip_\alpha+{1\over2}
p_\alpha p_{\alpha_1} { \partial\over  i\partial k_{\alpha_1}}
+{i\delta^2\over36}\times\right.
\nonumber\\
\biggl(5p_\alpha g_{\alpha_1\alpha_2}
&-&2p_{\alpha_2} g_{\alpha\alpha_1}\biggr)
\left.{ \partial\over  i\partial 
k_{\alpha_1}}{ \partial\over  i\partial k_{\alpha_2}}\right]
\delta^{(4)}(k)\;,
\nonumber\\
C_{\alpha\beta}(k,p)&=&-{(2\pi)^4\over24}{\langle {\bar q} q \rangle 
\over 3f_\pi}(p_\alpha g_{\beta\alpha_1}-p_\beta g_{\alpha \alpha_1})
\nonumber\\
&\times&\left[i{ \partial\over  i\partial k_{\alpha_1}}
-{p_{\alpha_2}\over2}
{ \partial\over  i\partial 
k_{\alpha_1}}{ \partial\over  i\partial k_{\alpha_2}}\right]
\delta^{(4)}(k)\;,
\label{ABC}
\eeqa
where $m_0^2$ and  $\delta^2$ are defined by
$\langle {\bar q} D^2 q \rangle = m_0^2
\langle {\bar q} q \rangle/2$,
$\langle 0| {\bar d} g_s {\tilde {\cal G}}^{\alpha\beta}
\gamma_\beta u | \pi(p) \rangle = i \delta^2 f_\pi p^\alpha$ ,
with ${\tilde {\cal G}}_{\alpha\beta}=\epsilon_{\alpha\beta\sigma\tau}
{\cal G}^{\sigma \tau}/2$ and ${\cal G}_{\alpha \beta} = 
t^A G_{\alpha \beta}$. 

The additional contributions to the OPE comes from the diagrams 
where one gluon, emitted from the $c$-quark propagator, is combined
with the quark-antiquark component. 
Taking the gluon stress tensor into the quark-antiquark component, one
can write down the correlation function into the form
\begin{eqnarray}
\Pi_{\mu} = 4 \int {d^4 k \over (2\pi)^4}
Tr \bigg[ \bigg(S_{\alpha\beta}(p_3-k)\gamma_\mu S(p_3-p_2-k)
\nonumber \\
+S(p_3-k)\gamma_\mu S_{\alpha\beta}(p_3-p_2-k)\bigg)\gamma_5 D^{\alpha
\beta}(k,p_1)\gamma_5\bigg]\;,
\label{corqqg}
\end{eqnarray}
where we have defined
\beq
S_{\alpha\beta}(k)=-{
\left [ k_\alpha \gamma_\beta - k_\beta \gamma_\alpha
+ (\fslash{k} + m_c) i \sigma_{\alpha\beta} \right ]\over 2 (k^2 -m^2_c)^2 }\ .
\eeq
The $c$-quark propagator with one gluon attached is given by \cite{rry}
$g_s{\cal G}_{\alpha \beta}S^{\alpha\beta}(k)$, and 
$$
D^{\alpha\beta}(k,p)= \gamma_5 \sigma_{\rho\lambda} E^{\rho\lambda\alpha
\beta}(k,p)
+\gamma^\tau \epsilon^{\alpha\beta\theta\delta}
F_{\tau\theta\delta}(k,p)\;.  
$$

Up to  twist-4 and at order ${\cal O} (p_\mu p_\nu)$, the two functions 
appearing above are given by \cite{kl,bbk}
\begin{eqnarray}
E^{\rho\lambda\alpha\beta} &=& {i\over32}\left[-{m_0^2 \langle 
{\bar q} q \rangle
\over  6 f_\pi}  (g^{\rho\alpha}g^{\lambda\beta}
-g^{\rho\beta}g^{\lambda\alpha})+f_{3\pi} [p^\alpha p^\rho g^{\lambda\beta}
\right.
\nonumber\\
&-&\left.
p^\beta p^\rho 
g^{\lambda\alpha}-p^\alpha p^\lambda g^{\rho\beta} +p^\beta p^\lambda 
g^{\rho\alpha}] \right](2\pi)^4\delta^{(4)}(k) 
\nonumber\\
F_{\tau\theta\delta} &=& -{i \delta^2 f_\pi \over 3 \times 32}
(p_\theta g_{\tau\delta} - p_\delta g_{\tau \theta})
(2\pi)^4 \delta^{(4)}(k) \ ,
\label{EF}
\end{eqnarray}
where $f_{3\pi}$ is defined by the vacuum-pion matrix element
$\langle 0| {\bar d} g_s \sigma_{\alpha\beta}\gamma_5{\tilde {\cal G}}^{\alpha
\beta}u | \pi(p) \rangle$ \cite{bbk}.

The phenomenological side of the correlation function, $\Pi_{\mu}$,
is obtained by the consideration of $J/\psi$, $\pi$, $D$ and $\bar{D}$ states 
contribution to the matrix element in Eq.~(\ref{cor}).
The hadronic amplitude is defined by the matrix element:
\beqa
i{\cal{M}}&=&\langle\psi(p_2,\mu)|~D(-p_3)~\bar{D}(-p_4)~\pi(p_1)
\rangle
\nonumber\\
&=&i~{\cal{M}}_{\mu}(p_1,p_2,p_3,p_4)~\epsilon_2^{\mu}\;.
\eeqa

The phenomenological side of the sum rule can be written as (for the
part of the hadronic amplitude that will contribute to the cross section)
\cite{nnmk}:
\beq
\Pi_{\mu}^{phen}=-{\mpsi f_\psi (\md^2 f_D/m_c)^2
~{\cal{M}}_{\mu}\over
(p_2^2-\mpsi^2)(p_3^2-\md^2)(p_4^2-\md^2)} + \mbox{h. r.}\; ,
\label{phendds}
\eeq
where h.~r. means higher resonances. The hadronic amplitude can be 
parametrized as:
\beq
{\cal{M}}_{\mu}=\Lambda~\epsilon_{\mu\alpha\beta\sigma}p_{1}^\alpha 
 p_3^\beta p_4^\sigma\,,
\label{strudd}
\eeq
where $\Lambda$ is 
the parameter that we will evaluate from the sum rules.

Inserting the results in Eqs.~(\ref{ABC}) and (\ref{EF}) into 
Eqs.~(\ref{corqq}) and (\ref{corqqg}) we can write a sum rule for the 
invariant structure appearing in Eq.~(\ref{strudd}). To improve the matching 
between
the phenomenological and theoretical sides we follow the usual procedure and
make a single Borel transformation to  all the external momenta taken to be 
equal: $-p_2^2=-p_3^2=-p_4^2=P^2\rightarrow M^2$. We get, in the approximation
$p_1<<p_2,p_3,p_4$: 
\beqa
{\Lambda+AM^2+BM^4\over\mpsi^2-\md^2}\Biggl[{e^{-\md^2/
M^2}\over M^2}- {e^{-\md^2/M^2}-
e^{-\mpsi^2/M^2} \over\mpsi^2-\md^2}\Biggr]
\nonumber\\
= {m_c^2\over\md^4\mpsi f_{D}^2f_\psi}
{e^{-m_c^2/M^2}\over M^2}  \Biggl[f_\pi-{2m_c\langle\bar{q}q\rangle\over
3f_\pi M^2}\;\;\;\;\;\;\;
\nonumber\\
-{f_\pi\delta^2\over18M^2}\Biggl(17+{5m_c^2\over M^2}\Biggr)\Biggr],
\;\;\;\;\;\;\;\;\;\;\;\;\;\;\;\;\;\;\;\;\,
\label{sr}
\eeqa
where we have transferred to the theoretical side the couplings of the currents
with the mesons. The problem of doing a 
single Borel transformation is the fact that terms 
associated with the pole-continuum  transitions are not
suppressed~\cite{io2}. In the present case we have two kinds of these 
transitions: double pole-continuum and single pole-continuum. In the limit
of similar meson masses it is easy to show that the Borel behavior of
the three-pole,  double pole-continuum and single pole-continuum
contributions are $e^{-m_M^2/M^2}/M^4,\,e^{-m_M^2/M^2}/M^2$ and
$e^{-m_M^2/M^2}$ respectively. Therefore, we can single out the three-pole
contribution from the others by introducing two parameters, $A$ and $B$,
in the 
phenomenological side of the sum rule, which will account for the 
double pole-continuum and single pole-continuum contributions respectively
\cite{kl,bnn,nos}.

The parameter values used in all calculations are  
$m_c=1.37\,\GeV$, $m_\pi=140\,\MeV$, $m_D=1.87\,\GeV$, $m_{D^*}=2.01\,
\GeV$, $\mpsi=3.097\,\GeV$, $f_\pi=131.5\,\MeV$, $f_\psi=270\,\MeV$,
$f_D=170\,\MeV$, $f_{D^*}=240\,\MeV$,
$\langle\overline{q}q\rangle=-(0.23)^3\,\GeV^3$, $m_0^2=0.8\,\GeV^2$,
$\delta^2=0.2\,\GeV^2$, $f_{3\pi}=0.0035\,\GeV^2$ \cite{bbk}.

\begin{figure}[htb]
\centerline{\psfig{figure=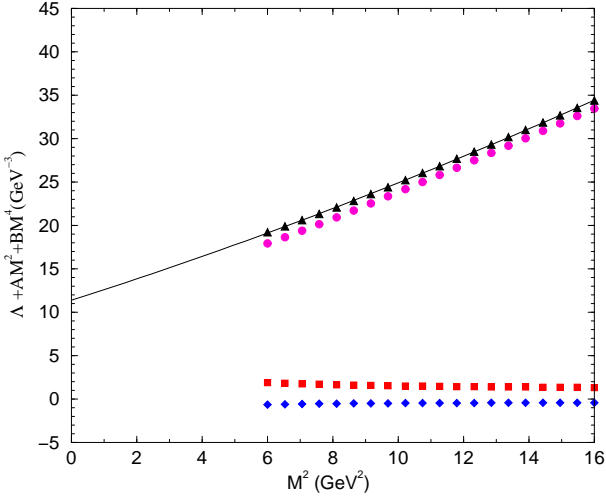,width=8cm,angle=0}}
\protect\caption{Amplitude of the process $J/\psi~\pi\rightarrow \bar{D}~D$
as a function of the Borel mass. The circles, squares and diamonds 
give the twist-2, 3 and 4 contributions to the sum rule. The triangles
give the result from  Eq.~(\ref{sr}). The solid line give the fit
to the QCDSR results.}
\label{fig1}
\end{figure}

In Fig.~1 we show the QCD sum rule results for 
$\Lambda+AM^2+BM^4$ as a function of $M^2$. The circles, squares and 
diamonds give the twist-2, 3 and 4 contributions respectively. The triangles
give the final QCDSR results. We see that the twist-3 and 4 contributions
are small as compared with the twist-2 contribution showing a ``convergence''
of the OPE expansion. The triangles
follow almost a straight line in the Borel region $6\leq M^2\leq16\,\GeV^2$,
indicating that the single pole-continuum transitions contribution is small.
The value of the amplitude $\Lambda$ is obtained by the extrapolation of
the fit to $M^2=0$~\cite{io2,bnn,nos}. Fitting the QCD sum rule results
to a quadratic form we get $\Lambda\simeq11.4\GeV^{-3}$.
As expected, in our approach $\Lambda$ is just a number and all dependence 
of ${\cal M}_{\mu}$ on particle momenta is contained 
in the Dirac structure. This is a consequence of our low energy
approximation.

Instead of using the experimental values for the meson decay constants, it
is also possible to use the respective sum rules, as done in \cite{nnmk}. 
The behavior of the results does not change significantly, leading only to a 
change in the value of the amplitude. Using the respective sum rules for 
the meson decay constants we get
$\Lambda\simeq14.9\GeV^{-3}$. We will use these two procedures to estimate
the errors in our calculation. Our results  agrees completely with
the value obtained in \cite{nnmk}.

The calculation of the sum rules for the processes $J/\psi~\pi\rightarrow 
\bar{D}~D^*$ and  $J/\psi~\pi\rightarrow \bar{D}^*~D^*$ can be done in a 
similar way. One has only to change the currents in Eq.~(\ref{cor}) by the
appropriate ones. The hadronic amplitudes for these two processes
can be written in terms of many different structures.
In terms of the structures that will contribute to the cross section we
can write
\begin{itemize} \item for the process $J/\psi~\pi\rightarrow \bar{D}~D^*$:
\beqa
{\cal{M}}_{\mu\nu}=\Lambda_1^{DD^*}p_{1\mu}p_{1\nu} + \Lambda_2^{DD^*}
p_{1\mu}p_{2\nu}
\nonumber\\
+\Lambda_3^{DD^*}p_{1\nu}p_{3\mu} + \Lambda_4^{DD^*} g_{\mu\nu}+ 
\Lambda_5^{DD^*}p_{2\nu}p_{3\mu}\;,
\label{strudds}
\eeqa
\item for the process $J/\psi~\pi\rightarrow \bar{D}^*~D^*$:
\beqa
&&{\cal{M}}_{\mu\nu\rho}=
\Lambda_1^{D^*D^*}~H_{\mu\nu\rho}+ \Lambda_2^{D^*D^*}~J_{\mu\nu\rho}
\nonumber\\
&+&\Lambda_3^{D^*D^*}g_{\nu\rho}\epsilon_{\mu\alpha\beta\gamma}p_{1}^\alpha 
 p_2^\beta p_3^\gamma + \Lambda_4^{D^*D^*}\epsilon_{\nu\rho\alpha\beta}
p_{3\mu} p_{1}^\alpha p_3^\beta 
\nonumber\\
&+& \Lambda_5^{D^*D^*}
\epsilon_{\nu\rho\alpha\beta}p_{3\mu} p_{1}^\alpha  p_2^\beta 
+\Lambda_6^{D^*D^*}\epsilon_{\mu\nu\alpha\beta}p_{3\rho}p_{1}^\alpha 
 p_2^\beta 
\nonumber\\
&+& \Lambda_7^{D^*D^*}\epsilon_{\mu\nu\alpha\beta}
p_{1\rho} p_{1}^\alpha 
 p_2^\beta + \Lambda_8^{D^*D^*}\epsilon_{\nu\rho\alpha\beta}p_{1\mu} 
p_{1}^\alpha  p_4^\beta 
\nonumber\\
&+&\Lambda_9^{D^*D^*}\epsilon_{\mu\nu\rho\alpha}p_{1}^\alpha 
+ \Lambda_{10}^{D^*D^*}\epsilon_{\nu\rho\alpha\beta}p_{1\mu} 
p_{1}^\alpha  p_3^\beta 
\nonumber\\
&+&\Lambda_{11}^{D^*D^*}\epsilon_{\mu\nu\rho\alpha}
p_{2}^\alpha 
+\Lambda_{12}^{D^*D^*}\epsilon_{\mu\nu\rho\alpha}
p_{3}^\alpha 
\nonumber\\
&+&\Lambda_{13}^{D^*D^*}\epsilon_{\mu\nu\alpha\beta}
p_{1\rho} p_{1}^\alpha  p_3^\beta + \Lambda_{14}^{D^*D^*}
\epsilon_{\mu\nu\alpha\beta}p_{3\rho} p_{1}^\alpha  p_3^\beta \;,
\label{strudsds}
\eeqa
\end{itemize} 
with
$H_{\mu\nu\rho}=(\epsilon_{\nu\alpha\beta\gamma}g_{\mu\rho}-\epsilon_{\rho
\alpha\beta\gamma}g_{\mu\nu})p_{1}^\alpha p_2^\beta p_3^\gamma + 
\epsilon_{\mu\rho\alpha\beta}p_{2\nu} p_1^\alpha p_2^\beta$ and
$J_{\mu\nu\rho}=(\epsilon_{\nu\rho\alpha\beta}p_{1\mu}+\epsilon_{\mu\rho
\alpha\beta}p_{1\nu}
+\epsilon_{\mu\nu\alpha\beta}p_{1\rho})p_{2}^\alpha p_3^\beta +
\epsilon_{\mu\nu\alpha\beta}p_{2\rho}p_{1}^\alpha p_3^\beta$. 
In principle
all the independent structures appearing in $H_{\mu\nu\rho}$ and 
$J_{\mu\nu\rho}$ would have independent parameters $\Lambda_i$. However,
since in our approach we get exactly the same sum rules for all of them,
we decided to group them with the same parameters.

The expressions for all 20 sum rules will be given elsewhere \cite{pre}.
 At this point it is important to stress that
the sum rule for the process $J/\psi~\pi\rightarrow \bar{D}~D$
is not particular, in general all the other sum rules are 
similar and contain  twist-2, twist-3 and twist-4 contributions corresponding
to the first, second, and third terms inside the brackets in the right hand 
side of Eq.~(\ref{sr}). Only the sum rules for $\Lambda_{10}^{D^*D^*}$ up
to $\Lambda_{14}^{D^*D^*}$ get only twist-4 contributions, and give results 
compatible  with zero. It is also interesting to notice that if we consider
only the twist-2 contributions we recover the sum rules obtained in 
ref.~\cite{nnmk}.

The results for all other sum rules show a similar behavior and the amplitude
can be extracted by the extrapolation of the fit to $M^2=0$. The values
for all the parameters are given in \cite{pre}.
In Eq.~(\ref{strudds}) the structures multiplying $\Lambda_4$ and $\Lambda_5$
break chiral symmetry \cite{nnr} and, therefore, will be neglected.

\begin{figure}[htb]
\centerline{\psfig{figure=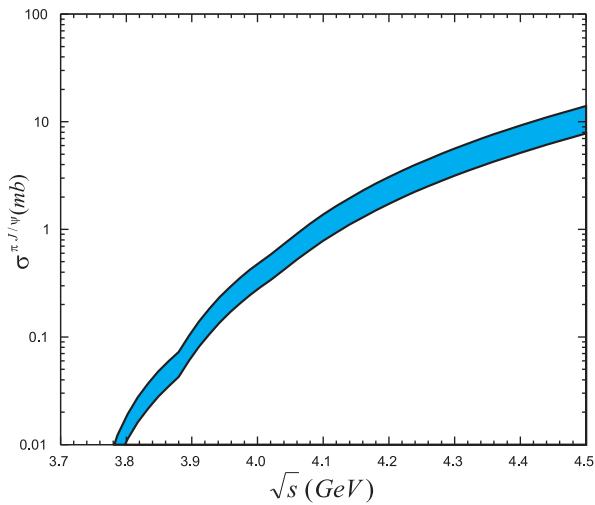,width=8cm,angle=0}}
\protect\caption{Total $J/\psi~\pi$ dissociation cross sections of the 
processes $J/\psi~\pi\rightarrow
\bar{D}~D^* + D~\bar{D}^* +\bar{D}~D+ \bar{D}^*~D^*$. The shaded area
give an evaluation of the uncertainties in our calculation.} 
\label{fig2}
\end{figure}

Having the QCD sum rule results for the amplitude of the three processes
$J/\psi~\pi\rightarrow \bar{D}~D^*,~\bar{D}~D,~\bar{D}^*~D^*$, 
we can evaluate  the cross section.
In Fig.~2 we show  the cross section 
for the $J/\psi-\pi$ dissociation. The shaded area give an evaluation of 
the uncertainties in our calculation
obtained with the two procedures described above.
It is important to keep in mind that,
since our sum rule was derived in the limit $p_1<<p_2,p_3,p_4$,
we can not extend our results to large values of $\sqrt{s}$. 

In a hadron gas, pions collide with the $J/\psi$ at different energies.
The momentum distribution of thermal pions in a hadron gas depends on the 
effective temperature $T$ with an approximate Bose-Einstein distribution.
Therefore, the relevant quantity is not the value of the cross section
at a given energy, but $\langle\sigma^{\pi J/\psi} v\rangle$ which is the 
product
of the dissociation cross section and the relative velocity averaged
over the energies of the pions.
\begin{figure}[htb]
\centerline{\psfig{figure=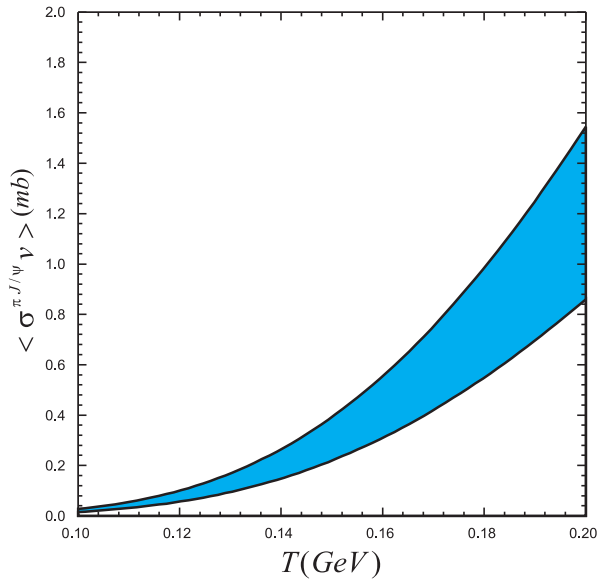,width=8cm,angle=0}}
\protect\caption{Thermal average of $J/\psi$ dissociation cross section by 
pions as a function of temperature $T$. The shaded area
give an evaluation of the uncertainties in our calculation.} 
\label{fig3}
\end{figure}

As shown in Fig.~3, $\langle\sigma^{\pi J/\psi} v\rangle$ increases with the
temperature. Since the $J/\psi$ dissociation by a pion requires energetic 
pions to overcome the energy threshold, it has a small thermal average
at low temperatures. The shaded area in Fig.~3
give an evaluation of the uncertainties in our calculation due to the two 
procedures used to extract the hadronic amplitudes. 

In conclusion, we have studied the $J/\psi$ dissociation cross section by pions
using the QCDSR technique, based on a three-point function using vacuum-pion 
correlation functions. We have estimated the hadronic amplitudes by working 
up to twist-4 in the limit $p_1<<p_2,p_3,p_4$. Our results are in agreement 
with
the former QCDSR calculation, done with a four-point function at the pion pole
\cite{nnmk}. Our results for the cross section as a function of $\sqrt{s}$
are smaller than the results using meson-exchange models (without form 
factors), but larger than the calculation based on quark-exchange models.

The dominant contribution to the hadronic amplitudes comes from the twist-2
operator, or equivalently, from the quark condensate. As we know that
the quark condensate is stronger in the vacuum and weaker in the interior
of hadrons, we can conclude that the charmonium ``sees'' and interacts
with the surface of the pions, where there is a ``halo'' of condensates.
This is way the cross section can be larger than the geometric value. In
our approach the continuous
growth of the cross section comes from the
growth of the phase space, as in the effective Lagrangian calculations.
In the short distance QCD calculation \cite{kha2,agga} the cross section 
also grows with $\sqrt{s}$, but the growth there is considerably
smaller because the rise in the gluon density cannot 
completely compensate the fall of the partonic cross section \cite{agga}.

The thermal average of the $J/\psi-\pi$ dissociation cross section increases
with the temperature and at $T=150\,\MeV$ we get
$\langle\sigma^{\pi J/\psi} v\rangle\sim0.2-0.4$ mb which is smaller than the
values used in phenomenological studies of $J/\psi$ absorption by comoving
hadrons in relativistic heavy ion collisions.

The same approach used here could be applied to calculate the $\Upsilon~\pi$
cross section. In a recent work \cite{agga} the $\sigma_{\Upsilon\pi}$
was computed using short distance QCD. Since we expect the non-perturbative
corrections to be less important for heavier systems, the differences
between short distance QCD and QCD sum rules should be smaller for
the $\Upsilon\pi$ system, and a systematic comparison between the two 
approaches becomes possible. We will address this point in the future.

Another possible extension of this work is the calculation of the $\chi_c~
\pi$ and $\eta_c~\pi$ cross sections. This can be done by replacing the
$j_\mu^{\psi}$ current in Eq.~(\ref{cor}) by the corresponding $\chi_c$
and $\eta_c$ currents. Unfortunately since $\psi'$ and $J/\psi$ have the 
same quantum numbers
and are, therefore, described by the same current, it is not possible,
in this approach, to estimate the $\psi'~\pi$ cross section. The $\psi'$ 
contribution to the present sum rule calculation is inside the parameters 
$A$ and $B$, in Eq.~(\ref{sr}) and cannot be separated from the other
higher mass states contributions.

\vspace{0.2cm}
We are grateful to J. H\"ufner and
H. Kim for fruitful discussions. M.N. would 
like to thank the hospitality and financial support from the Yonsei University
during her stay in Korea. This work was supported by CNPq and FAPESP-Brazil.



\begin{references}

\bibitem{ma86} T. Matsui and H. Satz, Phys. Lett. {\bf B178}, 416 (1986).
\bibitem{vo99} R. Vogt, Phys. Reports {\bf 310}, 197 (1999).
\bibitem{ge99} C. Gerschel and J. Huefner,
Ann. Rev. Nucl. Pat. Sci. {\bf49}, 255 (1999).

\bibitem{NA50} NA50 Collaboration (M.C. Abreu et al.), Phys. Lett.
{\bf B477}, 28 (2000).

\bibitem{kha} D. Kharzeev, C. Louren\c co, M. Nardi and H. Satz , 
Z. Phys. {\bf C74}, 307 (1997).

\bibitem{wong} C.Y. Wong,  Phys. Rev. Lett. {\bf 76}, 196 (1996).


\bibitem{cap} A. Capella et al., Phys. Lett. {\bf B393}, 431 
(1997).
%
\bibitem{bhp} G. Bhanot and M.E. Peskin, Nucl. Phys. {\bf B156}  
(1979) 391; M.E. Peskin, Nucl. Phys. {\bf B156}  (1979) 365.
%
\bibitem{kha2} D. Kharzeev and H. Satz, Phys. Lett. {\bf B334}  
(1994) 155.
%
\bibitem{agga} F. Arleo, P.-B. Gossiaux, T. Gousset and J. Aichelin,
Phys. Rev. {\bf D65} (2002) 014005.
%
\bibitem{dnn} H.G. Dosch,  F.S. Navarra,  M. Nielsen and M. Rueter, 
Phys. Lett. {\bf B466},  363 (1999). 
%
\bibitem{qmodel} Cheuk-Yin Wong, E. S. Swanson and T. Barnes, Phys. Rev.
{\bf C62} (2000) 045201; K. Martins, D. Blaschke and E. Quack, Phys. Rev.
{\bf C51} (1995) 2723.
%
\bibitem{mmodel} S.G. Matinyan and B. M\"uller, Phys. Rev. {\bf C58} 
(1998) 2994; Y. Oh, T. Song and S.H. Lee, Phys. Rev. {\bf C63} (2001)
034901; K.L. Haglin, Phys. Rev. {\bf C61} (2000) 031902;
K.L. Haglin and C. Gale, Phys. Rev. {\bf C63} (2001) 065201; Z. Lin and C.M. 
Ko, Phys. Rev. C62 (2000) 034903. 
%
\bibitem{nnr} F.S. Navarra, M. Nielsen and M.R. Robilotta, Phys. Rev. 
{\bf C64} (2001) 021901(R).
%
\bibitem{nnmk} F.S. Navarra, M. Nielsen, R.S. Marques de Carvalho and G. Krein,
Phys. Lett. {\bf B529} (2002) 87.
%
\bibitem{svz}  M.A. Shifman, A.I. Vainshtein and V.I. Zakharov, Nucl. 
Phys.  {\bf B120} (1977) 316.
%
\bibitem{rry} L.J. Reinders, H. Rubinstein and S. Yazaki, Phys.
Rep. {\bf 127} (1985) 1. 
%
\bibitem{kl}  H. Kim and S.H. Lee, Eur. Phys. Jour. {\bf C22} (2002) 707.
%
\bibitem{bbk}
V.M.~Belyaev, V.M.~Braun, A.~Khodjamirian and R.~Ruckl,
Phys.\ Rev.\ D {\bf 51} (1995) 6177.
%
\bibitem{io2}  B.L. Ioffe and A.V. Smilga, Nucl. Phys. {\bf B232} 109
(1984).
%
\bibitem{bnn} M.E. Bracco, F.S. Navarra and M. Nielsen, 
Phys. Lett. {\bf B454} (1999) 346.
%
\bibitem{nos} F.S. Navarra, M. Nielsen, M.E. Bracco, M. Chiapparini and
C.L. Schat, Phys. Lett.  {\bf B489},  319  (2000). 
%
\bibitem{pre} F.O. Dur\~aes et al., nucl-th/0211092.


\end{references}
\end{document}